\documentclass[12pt,cite]{article} 
\usepackage{epsfig}
\usepackage{subfigure}
\usepackage{amssymb}
\usepackage{multirow}
\usepackage{txfonts}
\begin{document}

\begin{center}
\noindent { \bf The Directional Dependence of Apertures, Limits and Sensitivity of the Lunar Cherenkov Technique to a UHE Neutrino Flux}\\[2em]
C.W. James\footnote{Corresponding author,clancy.james@adelaide.edu.au}, 
R.J. Protheroe\footnote{rprother@physics.adelaide.edu.au}\\
Department of Physics, The University of Adelaide,  Adelaide, SA 5005, Australia
\end{center}

\centerline{\bf Abstract} We use computer simulations to obtain the directional-dependence of the lunar Cherenkov technique for ultra-high energy (UHE) neutrino detection. We calculate the instantaneous effective area of past lunar Cherenkov experiments at Parkes, Goldstone (GLUE), and Kalyazin, as a function of neutrino arrival direction, finding that the potential sensitivity to a point source of UHE neutrinos for these experiments was as much as thirty times that to an isotropic flux, depending on the beam pointing-position and incident neutrino energy. Convolving our results with the known lunar positions during the Parkes and Goldstone experiments allows us to calculate an exposure map, and hence the directional-dependence of the combined limit imposed by these experiments. In the $10^{21}$--$10^{23}$~eV range, we find parts of the sky where the GLUE limit likely still dominates, and areas where none of the limits from either Parkes, GLUE, or experiments such as ANITA or FORTE are likely to be significant.  Hence a large anisotropic flux of UHE neutrinos from these regions is not yet excluded.

We also determine the directional dependence of the aperture of future planned experiments with ATCA, ASKAP and the SKA to a UHE neutrino flux, and calculate the potential annual exposure to astronomical objects as a function of angular distance from the lunar trajectory through celestial coordinates.  We find that the potential exposure of all experiments at $10^{20}$~eV and below, integrated over a calendar year, is flat out to $\sim 25^\circ$ from the lunar trajectory and then drops off rapidly.  The region of greater sensitivity includes much of the Supergalactic Plane, including M87 and Cen~A, as well as the Galactic Centre.  At higher energies this high-sensitivity region becomes broader, and we find that the potential exposure of the SKA at $10^{21}$~eV and above is almost uniform over celestial coordinates.

~

\section{Introduction}

In many respects, neutrinos are the perfect astronomical
messengers.  Since they interact rarely, any flux of neutrinos
will reach us almost unattenuated over cosmological distances.
Also, being uncharged, the paths of neutrinos will not be
perturbed in cosmic magnetic fields, and their arrival directions
will allow source identification. In particular, observations of
ultra-high energy (UHE) neutrinos are expected to help resolve
the origin of the UHE cosmic rays (CR), through measurements of
both the UHE neutrino energy spectrum and arrival directions, see
e.g.\ refs.\ \cite{Protheroe04,Falcke2004_SKAscienceCase}.
Cosmogenic UHE neutrinos are predicted to arise from the
interactions of the highest energy cosmic rays with background
photon fields.  Such interactions with photons of the cosmic
microwave background radiation (CMBR) were predicted by Greisen
\cite{Greisen} and by Zatsepin \& Kuzmin
\cite{Zatsepin_Kuzmin}  to cause a cut-off at $\sim 10^{20}$~eV in the
spectrum, the ``GZK cut-off''. UHE CR have
been observed with energies above $10^{20}$~eV 
\cite{Takeda03,Bird95,Connolly06,Abu-Zayyad05}, and measurements
of the UHE CR spectrum by the Pierre Auger Observatory are
consistent with the GZK cut-off \cite{Yamamoto07}. Therefore, a
flux of ``GZK neutrinos'' is almost guaranteed. Also, many models
of UHE CR production, such as the decay of super-massive dark
matter particles or topological defects, predict a flux of UHE
neutrinos, and limits on such a flux have already been used to rule out
$Z$-burst scenarios \cite{Gorham04,Barwick06} of UHE CR production.

The lunar Cherenkov technique is a method to detect UHE particles
(both cosmic rays and neutrinos) with ground-based radio
telescopes via the coherent Cherenkov radiation emitted upon
their interaction in the outer layers of the Moon
\cite{Dagkesamanskii}. The simulated sensitivity of the technique
is such that most models of the UHE neutrino flux are expected to
be readily detectable with the next generation of
radio-telescopes \cite{JamesProtheroe08}, in particular the
Square Kilometre Array (SKA) \cite{SKAwebsite}, while past
experiments at Parkes \cite{Hankins96, James07}, Goldstone
\cite{Gorham04}, and Kalyazin \cite{Beresnyak05} have already
placed significant limits on the UHE neutrino flux. However, the
dependence of such limits on particle arrival direction has yet
to be determined, despite the potential for using the arrival
directions of UHE neutrinos to point back to the source(s) of UHE
cosmic rays. In this paper, we use results of simulations to
analyse the directional properties of the lunar Cherenkov
technique with respect to UHE neutrino detection.

In Section \ref{thelunarcherenkovtechnique}, we introduce the
lunar Cherenkov technique, and explain why we expect a strong
directional dependence in sensitivity to UHE particles. We
calculate the instantaneous sensitivity of previous experiments
at Parkes, Goldstone, and Kalyazin in Section \ref{instantapp},
including the potential sensitivity to a point source. Using data
on the dates of observations at Parkes and Goldstone, in Section
\ref{past_limits} we calculate model-independent limits from
these experiments as a function of celestial coordinates. In
Section \ref{future_prospects} we present our calculations of the
directionally-dependent sensitivity of likely future observations
with the Australia Telescope Compact Array (ATCA), Australian SKA
Pathfinder (ASKAP), and the SKA, relative to potential sources of
UHE neutrinos.  Our simulation method was described fully in
our previous paper \cite{JamesProtheroe08} where we
re-evaluated limits from past lunar Cherenkov experiments, and
calculated the sensitivity of future experiments, to an isotropic
flux of UHE neutrinos.  We refer readers to this work for a
discussion of our method.

\section{Lunar Cherenkov Observations}
\label{thelunarcherenkovtechnique}

G.~A.~Askaryan~\cite{Askaryan} first noted an effect --- the
``Askaryan effect'' --- by which high-energy particles may be
detected remotely. Upon interacting in a dense material, high
energy particles will produce a cascade of secondary particles,
which will develop a negative charge excess by in-flight
annihilation of positrons and entrainment of electrons from the
surrounding material. At wavelengths larger than the apparent
dimensions of the shower as viewed by an observer, Cherenkov
radiation emitted by particles travelling faster than the speed
of light in that medium will add coherently. For a dense
dielectric medium (e.g.\ ice), this coherence condition typically
corresponds to frequencies of order of a few GHz or less at the
Cherenkov angle $\theta_C=\cos^{-1} (1/n)$ ($n$ the refractive
index), with the peak frequency reducing further from
$\theta_C$. Since the emitted power in the coherent regime scales
approximately with the square of both frequency and primary
particle energy, the resulting pulse of radiation from the
highest-energy particles rapidly becomes very strong, so that if
the interaction medium is transparent to radio waves, the emitted
pulse can readily escape the medium and be detected at large
distances. The Askaryan effect has now been experimentally
confirmed in sand \cite{Saltzberg_GorhamSAND01}, salt
\cite{Saltzberg_GorhamSALT05}, and ice
\cite{Saltzberg_GorhamICE07}, with measurements of the radiated
spectrum agreeing with theoretical predictions (e.g.\ ref.\
\cite{Munizetal02}).

Another medium in which to observe the Askaryan effect is the
lunar regolith, a sandy layer of ejecta covering the Moon to a
depth of $\sim$10~m. The regolith is known to have a low
radio-frequency attenuation, and it is likely that the
sub-regolith layers exhibit a similar property. As first proposed
by Dagkesamanskii and Zheleznykh~\cite{Dagkesamanskii}, and
attempted by Hankins, Ekers \& O'Sullivan~\cite{Hankins96} using
the Parkes radio telescope, observing the Moon with ground-based
radio-telescopes should allow the detection of the coherent
Cherenkov radiation from sufficiently high-energy particle
interactions in the outer lunar layers. The lunar Cherenkov
technique, as it is known, has subsequently been attempted at
Goldstone (GLUE) \cite{Gorham04} and Kalyazin \cite{Beresnyak05},
although the limits on an isotropic flux of UHE neutrinos from
these experiments have since been superseded by ANITA-lite, the
forerunner of the ANITA experiment, which aimed to detect the
Askaryan effect in the Antarctic ice sheet. However, simulation
results of lunar Cherenkov observations with the next generation
of radio-telescopes, such as ASKAP \cite{ASKAP07}, LOFAR (the
Low-Frequency Array) \cite{Scholten06}, and the SKA
\cite{JamesProtheroe08}, are very promising.  These instruments
are expected to provide a dramatic increase in the sensitivity of
the technique, allowing the detection (or elimination) of UHE
neutrino fluxes from most models of UHE CR production.  In the
meantime, the technique is currently being developed
experimentally by both the LUNASKA collaboration (our project),
utilising ATCA, and the NuMoon project, with the Westerbork
Synthesis Radio Telescope (WSRT) \cite{Scholten06}, and it has
also been the subject of several theoretical and Monte Carlo
studies
\cite{ZasHalzenStanev92,MunizZas01,GorhamRADHEP01,Beresnyak04}
together with our own recent work
\cite{JamesProtheroe08,James07}.

\begin{figure*}
\centerline{\epsfig{file=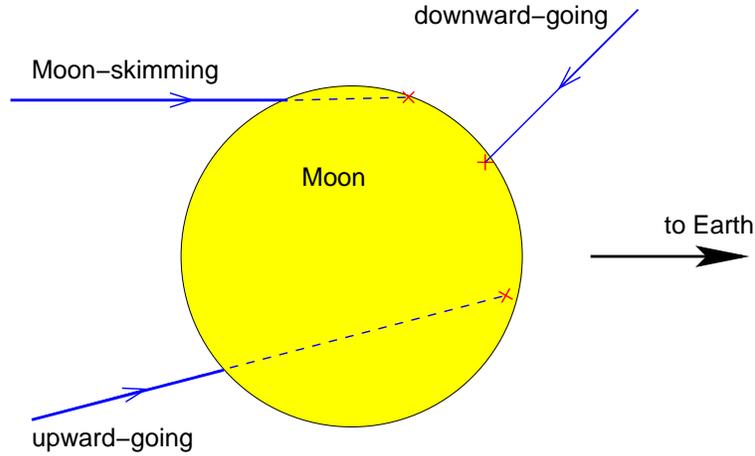, width=100mm} }
\caption{Interaction geometries of neutrinos.}
\label{InteractionGeometries}
\end{figure*}

\begin{figure*}
\centerline{\epsfig{file=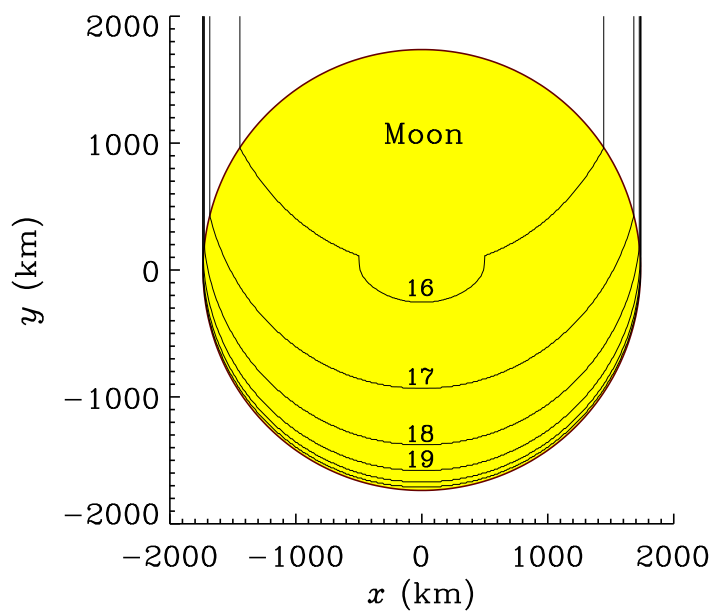, width=100mm}}
\caption{Locus of one neutrino mean free path (using cross-sections from Gandhi et al.\ \cite{Gandhi98}), as impact
parameter $x$ (with respect to the Moon's centre) varies from 0
to the lunar radius, and for energies $10^{16},10^{17}, \dots 10^{21}$~eV.  Neutrinos travel in the +$y$ direction,
i.e.\ upwards from bottom of figure.  Numbers attached to the curves
indicate $\log(E/$eV).}
\label{MoonTauContour_GQRS}
\end{figure*}

The probability of a lunar Cherenkov experiment detecting a given
particle incident on the Moon is expected to be highly dependent
upon the particle's arrival direction, due to a combination of
the opacity of the Moon to both UHE neutrinos and cosmic rays,
and the Cherenkov beam geometry --- at low frequencies the
Cherenkov cone is rather thick, but becomes thin at high
frequencies.  For the GLUE experiment, Gorham et al.\
\cite{GorhamRADHEP01} found that ``upcoming'' interactions (where the
direction of the primary just before interaction is towards the
local surface) are effectively ruled out, since these require the
particle to have penetrated through a large fraction of the lunar
bulk. Thus interactions are restricted to ``down-going'' (for both
CR and neutrinos) and ``Moon-skimming'' events (neutrinos only),
respectively where the primary interacts travelling into the
Moon, and where the primary interacts travelling nearly parallel
to the surface, having penetrated only a small portion of the
lunar limb.  Fig.~\ref{InteractionGeometries} illustrates such
interaction geometries, and in Fig.~\ref{MoonTauContour_GQRS} we
show the distance through the Moon to one neutrino interaction
length at various neutrino energies.  For skimming and
(especially for) down-going geometries, radiation at or near the
Cherenkov angle $\theta_C$ will be totally internally reflected
from the regolith-vacuum boundary, and only the weaker radiation
far from the Cherenkov angle will escape from the lunar
surface. As the Cherenkov cone narrows with increasing frequency,
only Moon-skimming and near Moon-skimming interactions (i.e.\
those with shallow incidence angles) will be detectable. Thus,
Beresnyak's finding \cite{Beresnyak04} that surface features with
favourably-aligned surface slopes --- enabling radiation nearer
$\theta_C$ to escape --- make a significant contribution to
experimental sensitivity.  Radiation close to the Cherenkov angle
is more likely to be refracted towards the direction of motion of
the primary particle than radiation far from the Cherenkov angle,
and so there should be a bias towards detecting radiation aligned
closely with the arrival direction of the primary particle.
Also, a radio telescope system with non-uniform sensitivity over
the Moon's visible surface will detect radiation coming from
different parts of the Moon with different efficiencies,
affecting the directional sensitivity of an experiment.

Previous simulations \cite{Gorham04, Beresnyak04} have shown
that, in combination, these effects --- especially for
high-frequency experiments --- lead to the majority of detectable
signals being expected to originate from the lunar limb. For past
experiments, where the individual beam size has been smaller than
the angular diameter of the Moon, this has made the beam pointing
position on the lunar surface critical in determining the
effective aperture. Future experiments with giant radio arrays
will likely face a comparable limit due to beam-forming
limitations over long baselines, necessitating a trade-off
between increasing the number of antenna used in real-time
triggering (and therefore sensitivity), and reducing the maximum
baseline over which to trigger in real time to increase coverage
of the lunar surface. A helpful measure in such an analysis will
be the effective aperture per unit solid angle as a function of
apparent position on the lunar surface.

\begin{figure*}
\centerline{\epsfig{file=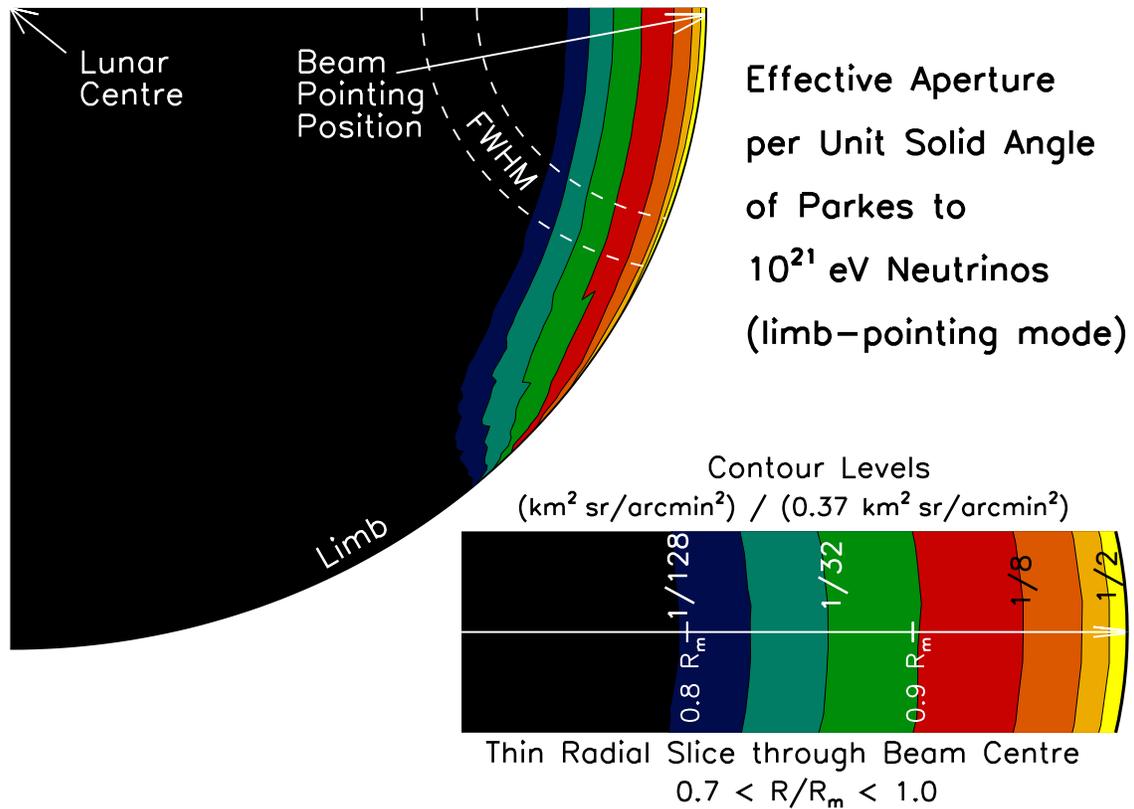, width=150mm}}
\caption{Effective aperture (km$^2$ sr), per arcmin$^2$ of lunar
disk, of the Parkes lunar Cherenkov experiment to $10^{21}$~eV
neutrinos as a function of the signal exit position on the 2-D
projection of the lunar surface.  Contour levels are
logarithmically-spaced fractions ($1/2,1/4,1/8$, etc.) of the
peak of $0.36$ km$^2$ sr per arcmin$^2$ of lunar disk.  Black
shading indicates a level of less than $1/128^{\rm th}$ of the
peak. We show only one quadrant of the Moon, since the plot is
vertically symmetric, and no signal is seen from the far side of
the Moon to the antenna beam. The finite range for the beam power
half-width half-maximum (FWHM) reflects the frequency ranges used
for triggering ($1.275$--$1.375$~GHz, and $1.475$--$1.575$~GHz).}
\label{origin}
\end{figure*}

Treating the Moon as its 2-D projection onto a (locally flat)
spherical shell of radius equal to the mean lunar orbital
distance ($3.844 \times 10^{8}$~m) about the Earth's centre, we
plot in Fig.\ \ref{origin} the effective aperture per arcmin$^2$
of lunar disk to $10^{21}$~eV neutrinos of the Parkes experiment
in limb-pointing mode as a function of the apparent origin of
detectable signals on the lunar surface. The very strong
limb-brightening is evident, as is the selection effect due to
the beam size. This is sufficient confirmation of the strong
geometrical dependence of the technique to motivate further
analysis.

While observations with the SKA promise to detect the observed
flux of UHE CR, the sensitivity is known only approximately,
since current methods of simulating large-scale lunar surface
roughness are inappropriate for modelling CR interactions
\cite{JamesProtheroe08}.  Unlike the case of UHE neutrinos, the
arrival directions of interacting CR will depend on the
local surface topography. This effect not only
reduces the total aperture to UHE CR, but will vary according to
CR arrival direction. We believe this extra degree of uncertainty
makes it inappropriate to simulate the directional sensitivity to
UHE CR with current modelling techniques, and so here we restrict
our directional analysis to UHE neutrinos.

\section{Directional Aperture}

\begin{figure*}
\centerline{\psfig{file=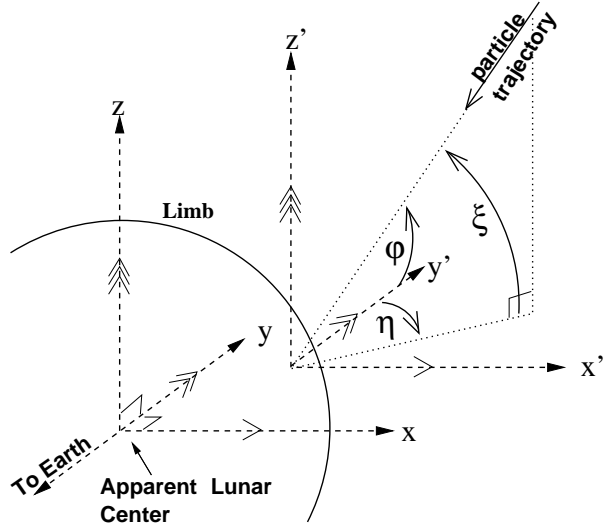, width=80mm}}
\caption{Coordinate definitions in the antenna-Moon system. The
Moon is treated as a 2-D projection onto the $x$--$z$ plane, with
the $y$-axis positive away from the observer (Earth). An incident
particle passing through some $(x'=0, \; y'=0,\; z'=0)$ (with
$\hat{x}' \parallel \hat{x},\; \hat{y}' \parallel \hat{y}$ and
$\hat{z}' \parallel \hat{z}$) has arrival direction defined by
angles $(\xi,\eta)$, with $\xi$ being the angle from the
$x'$--$y'$ plane ($-\pi/2 < \xi < \pi/2$), and $\eta$ the angle
between the $y'$-axis and the projection of the arrival direction
into the $x'-y'$ plane ($-\pi < \eta < \pi$).  We define $\varphi$ as the
angle between the Moon (i.e.\ the $\hat{y}$ direction) and the
particle's arrival direction, so that $\cos \varphi = \cos \xi
\cos \eta$.}
\label{xieta}
\end{figure*}

To place limits on an isotropic flux of high-energy neutrinos, an
experiment's effective aperture $A(E_{\nu})$ (km$^2$~sr) as a
function of particle energy $E_{\nu}$ is usually
calculated. Here, we treat the entire antenna-Moon system as our
detector, and calculate the acceptance as a function of both
$E_{\nu}$ \textit{and} arrival direction.  We calculate the
directional properties of $A(E_{\nu})$ via the experiment's
effective area $a(E_{\nu},\xi,\eta)$ (km$^2$), defined to be the
effective area to particles coming from a direction $(\xi,\eta)$,
specified relative to the antenna-Moon system as in Fig.\
\ref{xieta}. For the sake of brevity, we drop the explicit
dependence on energy and write $A \equiv A(E_{\nu})$, and
$a(\xi,\eta) \equiv a(E_{\nu},\xi,\eta)$.

In centre-pointing mode, the orientation of the $x$- and $z$-axes
about the $y$-axis is arbitrary, since the system is rotationally
symmetric about the $y$-axis.  In this case, $a(\xi,\eta)$ can be
written as $a(\varphi)$, where $\varphi$ is the angle between the
arrival direction and the apparent position of the Moon (see
Fig.\ \ref{xieta}).  When the antenna beam is not centred on the
Moon, we (arbitrarily) choose an orientation for the $x$- and $z$-axes such that the $x$-axis passes through the
beam centre. 

In the case of an isotropic flux, the effective aperture $A$ is
calculated via the product of the solid angle-averaged detection
probability $\bar{p}$ and the total lunar surface area of $4 \pi^2
R_m^2$. Similarly, the effective area $a (\xi,\eta)$ can be
calculated as the product of the detection probability
$p(\xi,\eta)$ and the lunar cross-sectional area, $\pi R_m^2$,
and can therefore be related to the effective aperture for an
isotropic flux by
\begin{eqnarray}
A & = & \oint a (\xi,\eta) \, d \Omega . \label{A_eq_int_area}
\end{eqnarray}
In randomising over arrival direction, our simulation has
effectively already performed this integration using Monte Carlo
methods, albeit in an indirect manner. To calculate $a(\xi,\eta)$
therefore, when we generate neutrinos of
energy $E_\nu$ incident on the Moon, we bin detectable events in
solid-angle bins, and from the resulting two-dimensional
histograms obtain the function $a(\xi,\eta)$. To speed up
computation, the simulation runs in a coordinate system in which
the particle's arrival direction is undefined, so we cannot force
the arrival directions to be evenly distributed in
$(\sin\xi,\eta)$, relying instead on Monte Carlo randomisation
for an even spread in solid angle. For a symmetric beam,
$a(\xi,\eta)=a(-\xi,\eta)$, which we use as a consistency check.

\subsection{Instantaneous Effective Area for Past Experiments}
\label{instantapp}

\begin{figure*}
\centerline{\epsfig{file=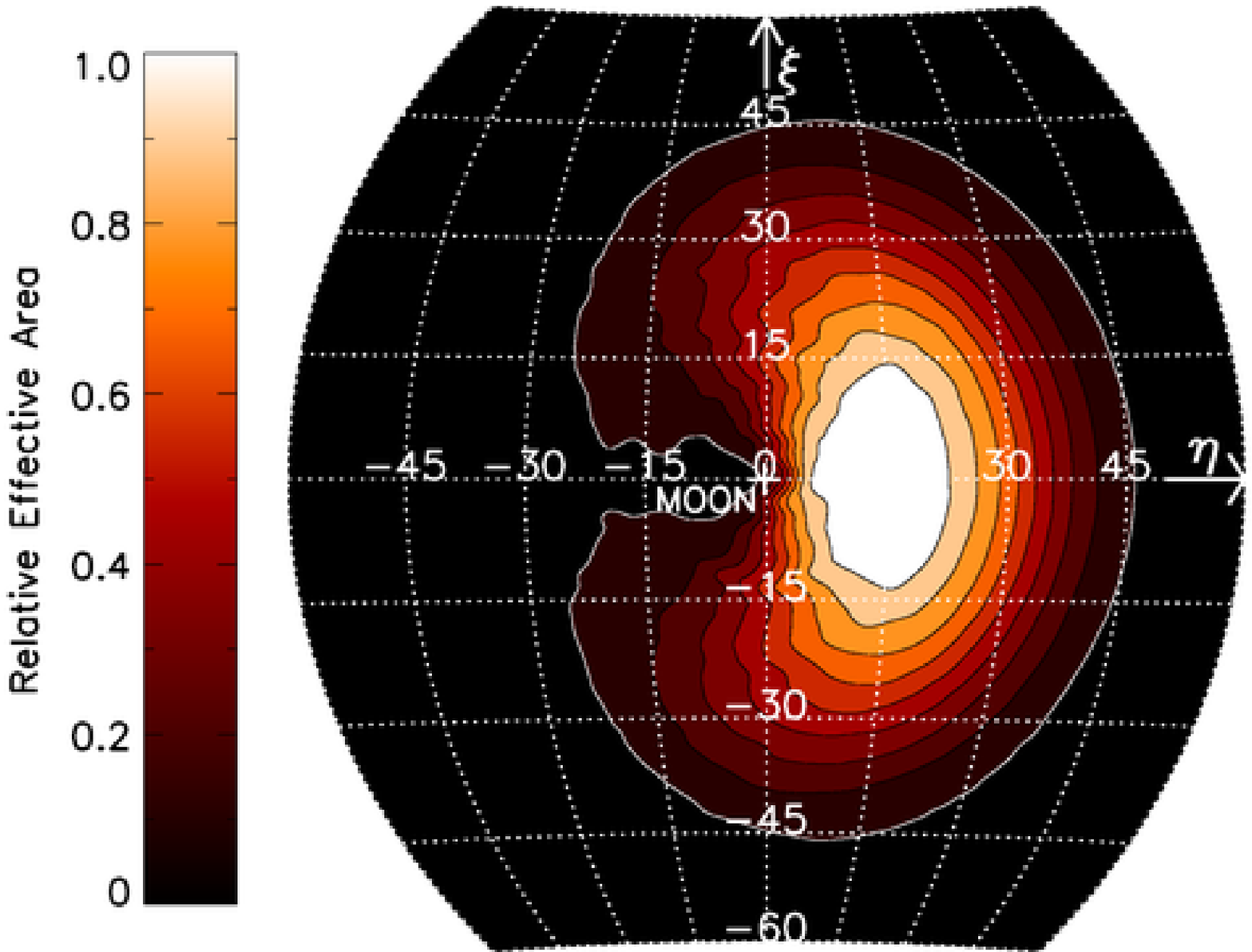, width=90mm} \epsfig{file=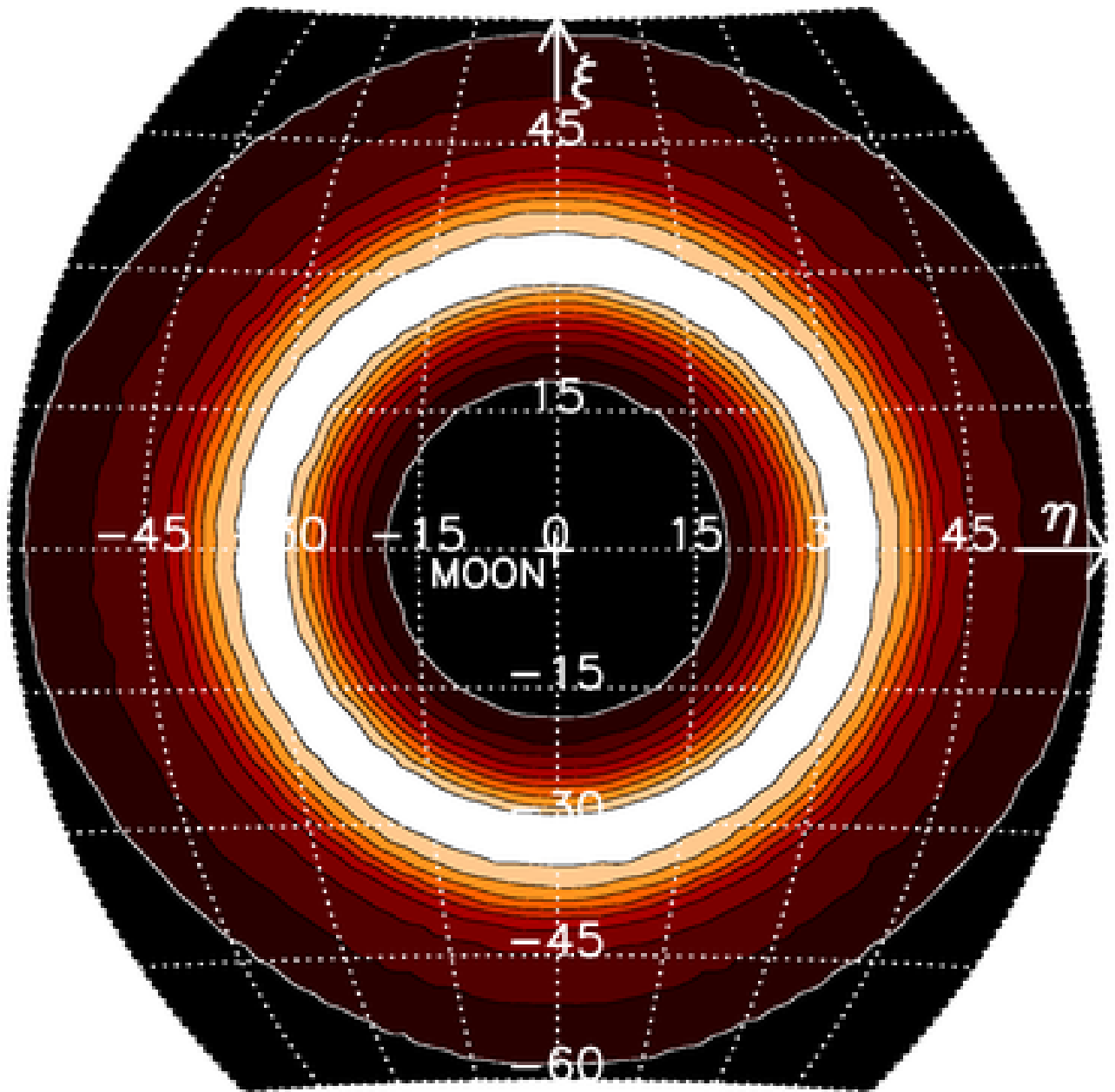,width=72mm}}
\caption{Normalised effective area $a(\xi,\eta)/a_{\rm max}$ of
the Parkes experiment to $10^{22}$~eV neutrinos, calculated
including a sub-regolith layer. Contours are at levels of $0.1$
$a_{\rm max}$, $0.2$ $a_{\rm max}$ \dots $0.9$ $a_{\rm max}$, with
shading corresponding to the upper value in each bin (e.g.\ white
is $0.9$-$1.0$ $a_{\rm max}$). Left: in limb-pointing mode, with
the beam at $(0^{\circ},0.25^{\circ})$; right: in centre-pointing
mode, with both the telescope beam and the Moon centred at
$(0^{\circ},0^{\circ})$. The peak values $a_{\rm max}$ are $81.5$
and $4.3$~km$^2$, for limb-pointing and centre-pointing, respectively.}
\label{parkes_instant}
\end{figure*}

We performed simulations to calculate $a(\xi,\eta)$ to neutrinos
for all configurations of previous experiments at Goldstone,
Kalyazin, and Parkes over a wide range of primary energies. We
find that the shape of $a(\xi,\eta)$ is similar for a given
configuration (e.g.\ limb-pointing) across all experiments, as
might be expected given the similarity of antenna sizes and
frequency ranges, and plot our estimates (calculated assuming the
presence of a sub-regolith layer with properties given in ref.\
\cite{JamesProtheroe08}) of $a(\xi,\eta)$ for $10^{22}$~eV
neutrinos in the limb-pointing and centre-pointing configurations
of the Parkes experiment, as shown in Fig.\ \ref{parkes_instant}.

Fig.\ \ref{parkes_instant} shows a characteristic ``kidney'' shape
of $a(\xi,\eta)$ in limb-pointing mode, and demonstrates the
symmetry $a(\xi,\eta) \equiv a(\varphi)$ in a centre-pointing
configuration. In centre-pointing mode, the directions with
highest sensitivity form an annular ring around the Moon with
peak effective area $a_{\rm max}$ occurring at $\varphi =
\varphi_{\rm max}$, with $\varphi_{\rm max} \approx 31^\circ$ at
$10^{22}$eV. The lowest sensitivity is to particles originating
from directions both too near (small $\varphi$) or far (large
$\varphi$) from the Moon's direction, respectively due to the
effective exclusion of upcoming interactions and the narrowness
of the Cherenkov cone. In limb-pointing configuration (beam
centred on $(\xi,\eta) = (0,0.25^{\circ})$) the peak effective
area $a_{\rm max}$ occurs at $(\xi,\eta) = (0,\eta_{\rm max})$,
with $\eta_{\rm max}\approx15^{\circ}$ at $10^{22}$eV, closer to
the Moon than in centre-pointing mode. Compared to the
centre-pointing configuration, the sensitivity to interaction
events in the targeted portion of the limb has increased, both
due to better beam reception and lower lunar thermal noise
levels, with a corresponding decrease in sensitivity to events
further along the limb.

In Table \ref{directionalitytbl} we give $a_{\rm max}$,
$\varphi_{\rm max}$ (centre-pointing mode) or $\eta_{\rm max}$
(limb-pointing mode), and the directionality ${\cal D}$ (to be
defined shortly) for previous lunar Cherenkov experiments at
$10^{21}$ eV, $10^{22}$ eV and $10^{23}$ eV. As particle energy
increases, radiation originating further from the Cherenkov
angle, and radiation reduced in intensity when refracted at large
angles, becomes detectable. 
The latter effect is the strongest, as evinced by the
position/locus of $a_{\rm max}$ moving closer to the Moon with
increasing neutrino energy.  The increased opacicty of the Moon
at high neutrino energies does not greatly influence the
sensitivity to different arrival directions since most of the
events are downgoing or Moon-skimming.
We also find that
in all cases, the spread of $a(\xi,\eta)$ about the peak is
greater at higher energies --- at lower energies,
$a(\xi,\eta)$ is non-zero only near the $(\xi,\eta)$
corresponding to $a_{\rm max}$.  The increased spreading is
somewhat off-set by the inclusion of the sub-regolith, which
tends to cause $a(\xi,\eta)$ to become more peaked at high
energies, since only for a small range of angles are interactions
in this layer detectable by high-frequency experiments.

The instantaneous effective area of the Parkes-Moon system covers
a relatively small part of the sky, although huge compared to the
antenna beam itself.  For an isotropic flux of UHE particles, of
course, the variation of $a(\xi,\eta)$ with $\xi$ and $\eta$ is
unimportant, as is the choice of observation times and pointing
position on the limb. Any anisotropy in the source of particles
will, however, make the event rate (or flux limit) dependent upon
the relative positions of the source(s), the Moon's centre, and
the antenna beam. There is thus significant scope for targeting
sources (suspected or discovered), as the greatest gain in
sensitivity is achieved for a point source lying in the direction
of maximum instantaneous effective area. In such a case, the
improvement in sensitivity over a blind observation is given by
the ratio of the peak $a_{\rm max}(E)$ to the
solid-angle-averaged value; we define this ratio to be the
directionality, ${\cal D}(E)$:
\begin{eqnarray}
{\cal D}(E) & = & {a_{\rm max}(E)} \left[ \frac{1}{4 \pi} \oint a (E,\xi,\eta) \, d \Omega \right]^{-1} \nonumber \\
& = & {4 \pi \, a_{\rm max}(E)}/{A(E)} .
\label{directionality}
\end{eqnarray}

The simulated values of ${\cal D}(E)$ for all three previous
experiments to UHE neutrinos, with the sub-regolith included, are
given in Table \ref{directionalitytbl}. As expected, ${\cal
D}(E)$ decreases with increasing primary particle energy, since
at higher energies a greater range of interaction geometries, and
hence arrival directions, are detectable.  Both $a_{\rm max}(E)$
and ${\cal D}(E)$ are larger in limb-pointing mode, since
limb-pointing increases sensitivity to a small range of arrival
directions at the expense of the majority. The high values
obtained for ${\cal D}(E)$, ranging from 7 to 28, indicate that
the limits set from these experiments are likely to be highly
anisotropic, depending strongly upon the observation times and
pointing positions of the antenna beam(s). Future experiments
should aim to choose parameters such as beam-pointing position
and observing schedule so that the peak sensitivity will be in
the direction of suspected (or, hopefully by then, discovered)
sources.

\subsection{Directional Limits from Parkes and GLUE}
\label{past_limits}

In the case of the Parkes experiment, almost all the limit arises
from just two hours spent pointing at the limb of the Moon spread
over two consecutive days -- unfortunately most of the
observation time was in centre-pointing mode for which the peak
effective area $a_{\rm max}$ was negligible (see Table
\ref{directionalitytbl}). Since the sensitivity is a function of
the Moon's position, which changes by approximately $13^{\circ}$
per day, the limit from the Parkes experiment will be
concentrated in a small patch of sky. Subsequent experiments at
Goldstone and Kalyazin spread observations over a longer period
of time, and are thus expected to produce a more evenly
distributed limit, although even a uniform spread of observations
over the lunar cycle will produce an anisotropic limit due to the
constraints of the Moon's orbit.

For an isotropic flux $I(E)$ (particles cm$^{-2}$ s$^{-1}$
sr$^{-1}$ GeV$^{-1}$), the expected event rate $N$ is given by
\begin{eqnarray}
N_{\rm obs} & = & t_{\rm obs} \int  A(E) I(E) dE .
\label{nobs1}
\end{eqnarray}
Hence, the usual method to place a limit on $I(E)$ is to
calculate the ``model-independent'' limit $I_{\rm lim}(E)$ as per
\cite{Lehtinen04}, i.e.\
\begin{eqnarray}
E I_{\rm lim} (E) & = & s_{\rm up} \left[  t_{\rm obs} \, A(E)  \right]^{-1} 
\label{isolim}
\end{eqnarray}
where $s_{\rm up}$ ($=2.3$ for a non-observation) reflects the
10\% confidence level for a Poisson distribution. For an
anisotropic flux $I(E,\alpha,\delta)$, the expected event rate
will be
\begin{eqnarray}
N_{\rm obs} & = &  \int dE  \int_{-1}^{1} d (\sin\delta) \int_{0}^{2 \pi}d \alpha \; I(E,\alpha,\delta) \, S(E,\alpha,\delta),
\label{nobs2}
\end{eqnarray}
where the experimental exposure $S$ is the effective area $a$ in
terms of celestial-coordinates ($\alpha,\delta$) integrated over
the observation time, i.e.\
\begin{eqnarray}
S(E,\alpha,\delta) & = & \int_{\rm obs} dt \, a (E,\alpha,\delta,t).
\label{calb}
\end{eqnarray}
The time-dependence of $a (E,\alpha,\delta,t)$ comes from
$(\alpha,\delta)$ being a time-dependent function of $(\xi,\eta)$
--- in which $a(E,\xi,\eta)$ is fixed --- due to the motion of
the Moon. Comparing Eqn.\ \ref{nobs2} with Eqn.\ \ref{nobs1}, the
directionally-dependent limit $F_{\rm lim} (E,\alpha,\delta)$
analogous to $I_{\rm lim} (E)$ should be calculated as in Eqn.\
\ref{directional_limit}, below,
\begin{eqnarray}
\label{directional_limit}
E F_{\rm lim} (E,\alpha,\delta) & = & s_{\rm up} \left[ S(E,\alpha,\delta) \right]^{-1}.
\end{eqnarray}
The simplest interpretation of $F_{\rm lim} (E,\alpha,\delta)$ is
as a limit on $F(E)$ (particles cm$^{-2}$ s$^{-1}$ GeV$^{-1}$)
from a point-source at celestial coordinates $(\alpha,\delta)$.

Calculating the exposure $S(E,\alpha,\delta)$ requires both
$a(E,\xi,\eta)$ (as calculated above) and an accurate record of
observation times and pointing positions. Having access to the
observation log, we were able to obtain such a record easily for
the Parkes experiment. For GLUE, we use the dates and duration of
observations in each configuration from Williams
\cite{Williams04}, assuming that the effective observation time
lay in a single block mid-way between moon-rise and moon-set on
each night. Sufficiently accurate times for observations at
Kalyazin could not be obtained.

\begin{figure*}
\centerline{\epsfig{file=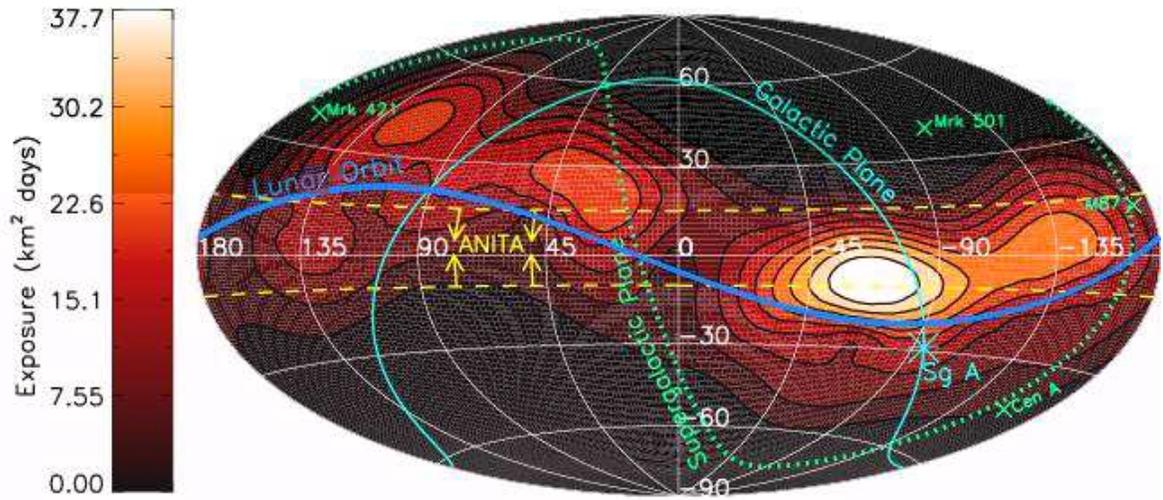, width=160mm}}
\caption{Combined exposure $S(E$=$10^{22}$~eV,~$\alpha,\delta)$
on a flux of $10^{22}$~eV neutrinos from experiments at Goldstone
and Parkes in J2000 coordinates. The dominant contribution is
from GLUE, due to the much longer observation time, with Parkes
contributing near $(\alpha,\delta) =
(135^{\circ},0^{\circ})$. Also shown is the declination
range $(-10^{\circ}<\delta<15^{\circ})$ of the ANITA experiment
\cite{Miocinovic05} --- the limit from ANITA-lite dominates in
this range. Contours are at 10\%, 20\% \dots 90\% of the peak
exposure of $37.7$~km$^2$ days.}
\label{glueparkes}
\end{figure*}

The combined exposure for the Parkes and GLUE experiments to
$10^{22}$~eV neutrinos is plotted in Fig.\ \ref{glueparkes},
obtained using discrete time-steps of $30$ minutes, and again
including the sub-regolith layer. The dominant contribution is
from GLUE, due to the much longer observation time, with Parkes
contributing near $(\alpha,\delta) =
(135^{\circ},0^{\circ})$. Interestingly, the spread of
observations is by no means uniform, with a peak exposure of
$37.7$~km$^2$ days at $(\alpha,\delta) \approx
(-64^{\circ},-8^{\circ})$. No suspected source of UHE particles
lies near this position. While the declination range ($-10^\circ
< \delta < 15^\circ$; also plotted) corresponding to high
sensitivity observations by ANITA-lite includes this direction,
the GLUE limit will be stronger away from this range. A complete
plot for experiments with significant limits at $10^{22}$~eV
would also include FORTE and Kalyazin, both of which should also
have highly anisotropic exposures, which might be expected to be
qualitatively similar to ANITA-lite and GLUE
respectively. Therefore, from a purely observational point of
view, there is scope for a potentially large flux of $E \approx
10^{22}$~eV neutrinos originating from parts of the sky to which
the accumulated exposure of all experiments is negligible.

\section{Potential Exposure of Future Experiments}
\label{future_prospects}

Future UHE neutrino experiments are likely to operate at lower
frequencies with smaller dishes than past experiments, for which
the maximum aperture will be obtained in a centre-pointing
configuration. The effective area, $a(E,\xi,\eta)$, will have a
characteristic annular shape, albeit somewhat broader to reflect
the lower observation frequencies. While the shape of
$a(E,\xi,\eta)$ shown in Fig.\ \ref{parkes_instant} reflects the
exposure of past experiments, where observation times were short
and sporadic, any serious future effort should involve
observations spread over a large time period. In such a case, the
potential exposure will approximately depend only on the angular
distance from the lunar orbital plane --- ``approximately'' because
the lunar orbit is not circular, nor will the Moon's visibility
be uniform over the orbit. Over a typical experimental lifetime,
an object's position with respect to the lunar orbital plane will
vary by $\sim \pm5^\circ$, since $5^{\circ}$ is approximately the
inclination of the lunar orbit to the ecliptic, with nodal
precession period of 18.6 years. Thus a future experiment's
potential exposure function will measure its ability to detect
UHE particles coming from astrophysical objects at various
angular distances from the lunar orbital plane.

\begin{figure*}
\centerline{\epsfig{file=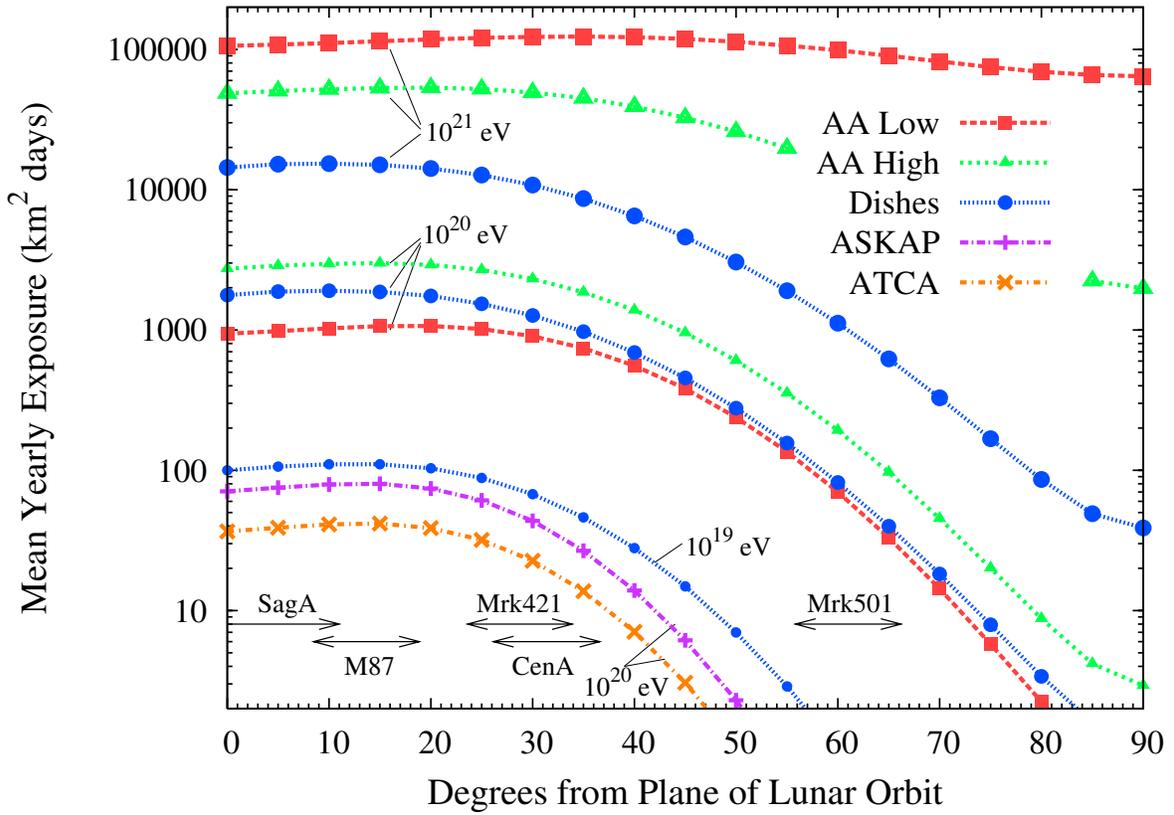, width=160mm}}
\caption{Potential exposure (km$^2$ days) for a calendar year of
future lunar Cherenkov experiments to neutrinos at specified
energies as a function of angular distance from the apparent
plane of the Moon's orbit. The range in apparent angular
distances from the Moon's orbit to astronomical objects reflects
the precession of the lunar orbital nodes ($18.6$ year period) in
the plane of the ecliptic -- differences in angular distance to
the apparent lunar orbit between experiments at different
latitudes arising due to parallax are negligible in comparison.}
\label{exposure}
\end{figure*}

Fig.\ \ref{exposure} plots the potential exposure from a calendar year's
equivalent observations of future experiments under the
aforementioned assumptions, weighted by mean lunar visibility, and
calculated with the sub-regolith included. When integrated over a
lunar cycle, the potential exposure function is almost flat
within $25^\circ$ of the lunar orbit for all instruments and at
all energies, dropping rapidly at large angular distances for all
but the low-frequency AA at $10^{21}$~eV. The greatest effect on
the shape of the exposure function at a given neutrino energy is
the sensitivity of the experiment to those particles --- as the
total exposure increases, the coverage broadens also.

For a given total exposure, experiments observing at lower
frequencies have a broader coverage, due to the width of the
Cherenkov cone increasing with decreasing frequency. However,
this effect is negligible in the case of the three SKA frequency
ranges until $10^{21}$~eV, where only the coverage of the
low-frequency AA becomes almost uniform. This is because the detection
threshold for the high-f AA and the dishes is lower.

Importantly for a GZK neutrino flux, the SKA dishes will have
greatly reduced exposure to any $\sim10^{19}$~eV neutrinos
arriving from further than $30^{\circ}$ from the lunar orbital
plane, representing half the sky. Viewed another way, the
potential exposure within $\sim 30^{\circ}$ of the plane of the
lunar orbit will be almost twice that of the averaged value.
Serendipitously, the ``phase'' of the ecliptic --- and hence lunar
orbit --- in right ascension (see Fig.\ \ref{glueparkes}) about
the celestial equator is nearly matched to that of the
Supergalactic Plane, from which an excess of UHE particles might
be detected, and objects of interest such as M87 and Cen A will
be readily visible. However, a large fraction of the sky, centred
at the North and South ecliptic poles (NEP and SEP) and including
(for example) Mrk 501, will remain inaccessible to the lunar
Cherenkov technique for neutrinos below $10^{21}$~eV. ANITA and
any follow-up experiments will be unlikely to have significant
exposure far from the celestial equator.  Also, since at energies much
above $10^{19}$~eV the apertures of IceCUBE, Auger and others to
UHE neutrinos are relatively low, it is unlikely that this UHE
neutrino energy/arrival direction parameter space, i.e.\ above
$\sim10^{19}$~eV and near the celestial poles, will be probed by
any current or near-future instruments.

\section{Conclusions}

Our results show that current limits on an $E_{\nu} \gtrsim
10^{20}$~eV neutrino flux are highly anisotropic, and therefore
that there is scope for a potentially large flux of UHE neutrinos
from arrival directions to which current limits are negligible.
The importance of this result depends upon the degree of
anisotropy in the UHE neutrino flux.  Our view is that since UHE
neutrino observations have the potential to probe some of the
most exotic phenomena in the universe, at energies far beyond
that tested in terrestrial laboratories, we should take seriously
the fact that no existing or currently planned experiment has
excluded or will be able to exclude high fluxes of UHE neutrinos
from large regions of sky near the celestial poles.  However, if
in the future it will be possible to use the low-frequency
aperture array, anticipated to be one of the key components of the
SKA, for lunar Cherenkov work the whole sky will be
accessible to $>10^{21}$~eV neutrino observations.

In conclusion, we have shown that future lunar Cherenkov
observations with ATCA, ASKAP, and the SKA will be able to detect
UHE neutrinos arriving from a broad range of directions,
including those coming from the direction of objects of interest
such as Cen A and M87.  The exposure near the lunar orbit is
high, and disproportionately strong limits (event rates) on
potential (discovered) sources could be placed via a careful
choice of observation time.  Experiments such as ANITA
view declinations $-10^{\circ}<\delta<15^{\circ}$ whereas
lunar Cherenkov experiments typically view a band within $\sim
30^\circ$ of the plane of the lunar orbit, or an even broader
band becoming almost isotropic at low frequencies at $10^{21}$~eV
and above.  Thus the lunar Cherenkov technique complements others
in that it covers significant parts of the sky inaccessible to
experiments such as ANITA.  It also offers the possibility of
very large effective apertures over a wide range of energies
above $\sim 10^{19}$~eV depending on frequency band.

\section*{Acknowledgments}
We thank J.~Alvarez-Mu\~{n}iz for his numerous suggestions
and helpful advice. This research was supported under the Australian
Research Council's Discovery Project funding scheme (project
number DP0559991).

\newpage

\begin{table}
\begin{center}
\begin{tabular}{ | l l | c c c | c c c  | c c c |}
\hline
& & \multicolumn{3}{ |c}{$10^{21}$~eV} & \multicolumn{3}{|c}{$10^{22}$~eV} & \multicolumn{3}{|c |}{$10^{23}$~eV} \\
& & $a_{\rm max}$ & $\theta_{\rm max}$  & ${\cal D}$ & $a_{\rm max}$  & $\theta_{\rm max}$  & ${\cal D}$ & $a_{\rm max}$ & $\theta_{\rm max}$ & ${\cal D}$ \\
\hline
\hline
Parkes & Limb &	5.03 &	16.2 &	27 &  		81.5 &	15.0 &	19 &		580 &	14.4 &	14 \\
%& Centre &	0.27 &	55 &	56 &		4.4 &	31.9 &	10.8 &		66.4 &	21.7 &	10.1 \\
& Centre &	0.0035 & 42.0 &	28 &		4.3 &	31.2 &	10 &		106 &	26.4 &	7 \\
\hline
GLUE & Limb&	2.11 &	18 &	27 &		32 &	15.6 &	20 &		232 &	16.2 &	15 \\
& Half-limb &	1.12 &	20 &	26 &		32.2 &	16 &	16 &		242 &	15.6 &	12 \\
& Centre &	0.02 &	39 &	26 &		6.5 &	27.6 &	10 &		105 &	19.2 &	8 \\
\hline
Kalyazin & Limb & 1.26 & 16.8 &	26 &		23.3 &	16.8 &	21 &		190 &	16.2 &	15 \\
\hline
\end{tabular}
\end{center}
\caption{Directional properties of past lunar Cherenkov
experiments for UHE neutrinos: the maximum effective area,
$a_{\rm max}$ (km$^2$); angle from the lunar centre of peak
sensitivity, $\theta_{\rm max} \;(^\circ )$ where $\theta_{\rm max}\equiv\varphi_{\rm max}$ (centre-pointing mode) and $\theta_{\rm max}\equiv\eta_{\rm max}$ (limb and half-limb pointing mode); and the
directionality, ${\cal D}$ (see Eqn.\ \ref{directionality}),
calculated with the sub-regolith layer included.}
\label{directionalitytbl}
\end{table}

\end{document}